\documentclass[%
 reprint,
superscriptaddress,
 amsmath,amssymb,
 prl
]{revtex4-2}

\usepackage{braket}
\usepackage{graphicx}
\usepackage{dcolumn}
\usepackage{bm}
\usepackage{hyperref}

\begin{document}

\preprint{APS}

\title{The Hidden Phase of the Spin-Boson Model}

\author{Florian Otterpohl}
\affiliation{%
Center for Computational Quantum Physics, Flatiron Institute, New York, New York 10010,
USA}%
\affiliation{%
I. Institut f\"ur Theoretische Physik, Universit\"at Hamburg, Notkestr. 9, 22607 Hamburg, Germany\\
}%
\author{Peter Nalbach}
\affiliation{
 Fachbereich Wirtschaft \& Informationstechnik, Westf\"alische Hochschule, M\"unsterstr. 265 46397 Bocholt, Germany
}%
\author{Michael Thorwart}%
\affiliation{%
 I. Institut f\"ur Theoretische Physik, Universit\"at Hamburg, Notkestr. 9, 22607 Hamburg, Germany\\
}%

\date{\today}

\begin{abstract}
A quantum two-level system immersed in a sub-Ohmic bath experiences enhanced low-frequency quantum statistical fluctuations which render the nonequilibrium quantum dynamics highly non-Markovian. Upon using the numerically exact time-evolving matrix product operator approach, we investigate the phase diagram of the polarization dynamics. In addition to the known phases of damped coherent oscillatory dynamics and overdamped decay, we identify a new third region in the phase diagram for strong coupling showing an aperiodic behaviour. We determine the corresponding phase boundaries. The dynamics of the quantum two-state system herein is not coherent by itself but slaved to the oscillatory bath dynamics.
\end{abstract}

\maketitle

%

\textit{Introduction}. ---%
Dissipative environments are the cause of relaxation and decoherence in quantum systems. Accordingly, understanding and modelling their influence and subsequently tailoring their impact is relevant to many research areas \cite{leggett_dynamics_1987,weiss_quantum_2012}. At strong system-bath coupling, dissipative environments may also lead to fully incoherent dynamics or even complete suppression of the dynamics (localization). Generally it is believed that, while prominent environmental modes may well cause coherent oscillations in a quantum system, a broadband reservoir destroys coherence at sufficiently strong coupling \cite{iles-smith_environmental_2014,maguire_environmental_2019}.

Typically, a two-state system interacting with harmonic degrees of freedom (spin-boson model) is studied \cite{leggett_dynamics_1987,weiss_quantum_2012}. At low temperatures and weak coupling damped coherent oscillations are observed while at stronger dissipation a classical incoherent decay towards thermal equilibrium is found. For an Ohmic bath with spectrum $J(\omega)\propto \alpha \omega^s$ and spectral exponent $s=1$ the ratio of damping rate and oscillation frequency is independent of the oscillation frequency. With increasing coupling $\alpha$ the dynamics turns incoherent and, for even larger coupling, a zero temperature phase transition towards a localized phase is observed \cite{leggett_dynamics_1987,weiss_quantum_2012}. 

While super-Ohmic spectra with $s>1$ with more pronounced high-frequency modes do not show incoherent dynamics nor a localization transition, reservoirs with more pronounced low frequency spectra such as sub-Ohmic ones with $s<1$ exhibit a localization phase transition  \cite{anders_equilibrium_2007, winter_quantum_2009, alvermann_sparse_2009} and are relevant for quantum impurity systems \cite{si_locally_2001,gegenwart_multiple_2007}. 
In contrast, the dynamic transition to incoherent behaviour is far less studied although this class of reservoirs constitutes the dominant noise source in, for example, nanomechanical oscillators systems \cite{groblacher_observation_2015} and superconducting qubit architectures (specifically charge noise generated by two-level fluctuators) \cite{AstafievPRL2004, ShnirmanPRL2005, YouPRR2021}. Although originally believed to be always incoherent \cite{leggett_dynamics_1987}, it is now well established that for $1/2 \lesssim s\le 1$ with increasing coupling the dynamics turns from damped oscillatory to incoherent, and for larger coupling a transition to a localized phase takes place \cite{chin_coherent-incoherent_2006,nalbach_ultraslow_2010, kast_persistence_2013, duan_zero-temperature_2017}. For $0\le s\lesssim 1/2$, however, Kast and Ankerhold \cite{kast_persistence_2013} showed that oscillatory dynamics persists for arbitrary coupling strength for an initially polarized bath. Pure dephasing sub-Ohmic reservoirs influence a quantum two-state system in a similar way and result in overdamping only for $\omega_c\rightarrow \infty$ and $s\gtrsim 0.1$ \cite{nalbach_crossover_2013}.

This leaves two important questions: What happens when crossing the supposed phase boundary at $s\simeq 1/2$ for a fixed but strong coupling and why can strong sub-Ohmic fluctuations not turn the system dynamics incoherent?
We investigate for $1/2\lesssim s \le 1$ and increasing coupling to the environment the dynamics which turns from coherent to incoherent for a coupling $\alpha_D(s)$ which depends on the spectral exponent $s$. For even larger couplings we encounter a sharp phase boundary at a coupling $\alpha_B(s)$ to aperiodic behaviour, i.e., the dynamics is not fully incoherent but exhibits a single turnaround. This new phase extends to all $0\le s \le 1$. 
For $0\le s \lesssim 1/2$, where Ref.~\cite{kast_persistence_2013} reports damped oscillatory dynamics for all couplings, we find with increasing coupling a sharp transition from oscillatory dynamics with various minima and maxima to aperiodic dynamics with a single minimum for an initially unpolarized bath. The frequency related to the 
minimum is proportional to the reservoir cut-off frequency. Accordingly, this dynamics is not generated by the central quantum system but by the reservoir and it turns incoherent for $\omega_c\rightarrow\infty$.
We simulate the dynamics in a broad range of parameters in order to map the new phase diagram of the dynamical behaviour of the spin-boson model.

\textit{Model}. ---%
We consider the symmetric spin-boson model ($\hbar = 1$, $k_B = 1$) with the Hamiltonian
\begin{eqnarray}
\hat{H} &&= \hat{H}_S + \hat{H}_B + \hat{H}_{\text{int}} \nonumber \\
&&=\frac{\Omega}{2} \hat{\sigma}_x + \frac{1}{2} \sum_j \left( \hat{p}_j^2 + \omega_j\hat{x}_j^2 \right)  + \frac{\hat{\sigma}_z}{2} \hat{\xi},
\end{eqnarray}
where $\hat{\xi} = \sum_j c_j \hat{x}_j$, $\Omega$ is the tunneling splitting, and $\hat{\sigma}_{x/z}$ are the Pauli matrices. The bath consists of harmonic oscillators with momenta $\hat{p}_j$, angular frequencies $\omega_j$, and positions $\hat{x}_j$ which are coupled via coupling constants $c_j$ to the spin. The bath has the spectral density
\begin{equation}
    J(\omega ) = \sum_j \frac{c_j^2}{2 \omega_j} \delta (\omega - \omega_j ) = 2 \alpha \frac{\omega^s}{\omega_c^{s-1}} e^{- \omega / \omega_c} ,
\end{equation}
with the high-frequency cut-off  $\omega_c = 10 \Omega$ (unless specified otherwise) and the coupling strength $\alpha$. We focus on the (sub-)Ohmic regime where $0 \leq s \leq 1$. \\
We calculate the time-dependent polarization $P(t) = \Braket{\sigma_z} (t)$, using a factorizing initial preparation of the system with $P(0) = 1$ and the thermal distribution of the initially uncoupled (i.e., unpolarized) bath at zero temperature.

\textit{Method}. --- %
To determine $P(t)$ we use the numerically exact real-time quasiadiabatic propagator path integral (QUAPI) \cite{makri_numerical_1995,makri_tensor_1995,makri_tensor_1995-1,makri_small_2020-1,strathearn_efficient_2018,strathearn_modelling_2020}. Once the bath oscillators have been integrated out, an effective dynamics of the system arises which is nonlocal in time. To treat the highly entangled system-bath dynamics of the sub-Ohmic spin-boson model, we make use of the time-evolving matrix product operator (TEMPO) technique \cite{strathearn_efficient_2018,strathearn_modelling_2020}, see Supplemental Material for further details. It allows us to avoid the conventional QUAPI memory cutoff as it rewrites the arising discrete path-sum in terms of a numerically highly efficient tensor network.

\begin{figure}[t]
\includegraphics[width=86mm]{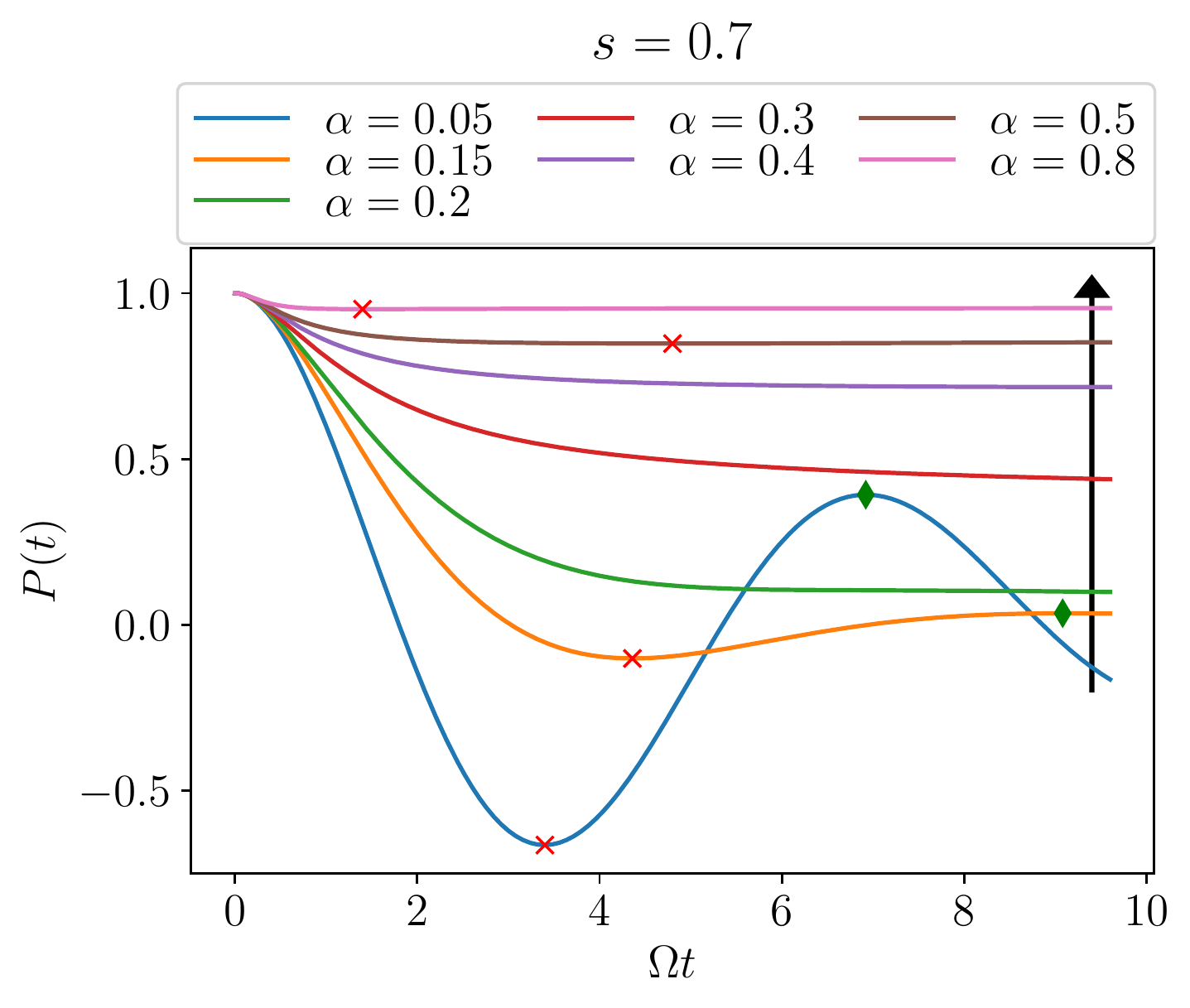}
\caption{Polarization $P(t)$ for $T=0$, $s = 0.7$, and different couplings strengths $\alpha$. The arrow intersects the lines in ascending order of coupling strengths. Local minima (maxima) are marked by a red cross (green diamond).}
\label{sectionplot_s_0.7}
\end{figure}

\textit{Polarization dynamics}. --- %
At first, we investigate the dynamics for a fixed $s=0.7$ for increasing coupling $\alpha$. Fig.~\ref{sectionplot_s_0.7} shows the time-dependent polarization $P(t)$ for times up to $\Omega t\lesssim 10$. For $\alpha = 0.05$ and $\alpha = 0.15 $, we observe damped oscillatory or coherent dynamics with a minimum (marked by a red cross in Fig.~\ref{sectionplot_s_0.7}) and a maximum (marked by a green diamond) visible in the studied time frame. But for couplings $\alpha = 0.2 $ and $\alpha = 0.4$, we find a monotonic decay highlighting purely incoherent dynamics. Surprisingly, for larger couplings $\alpha \ge 0.5$ we find again a minimum but no subsequent maximum. We denote this dynamics pseudo-coherent in the following.

\begin{figure}[t]
\includegraphics[width=86mm]{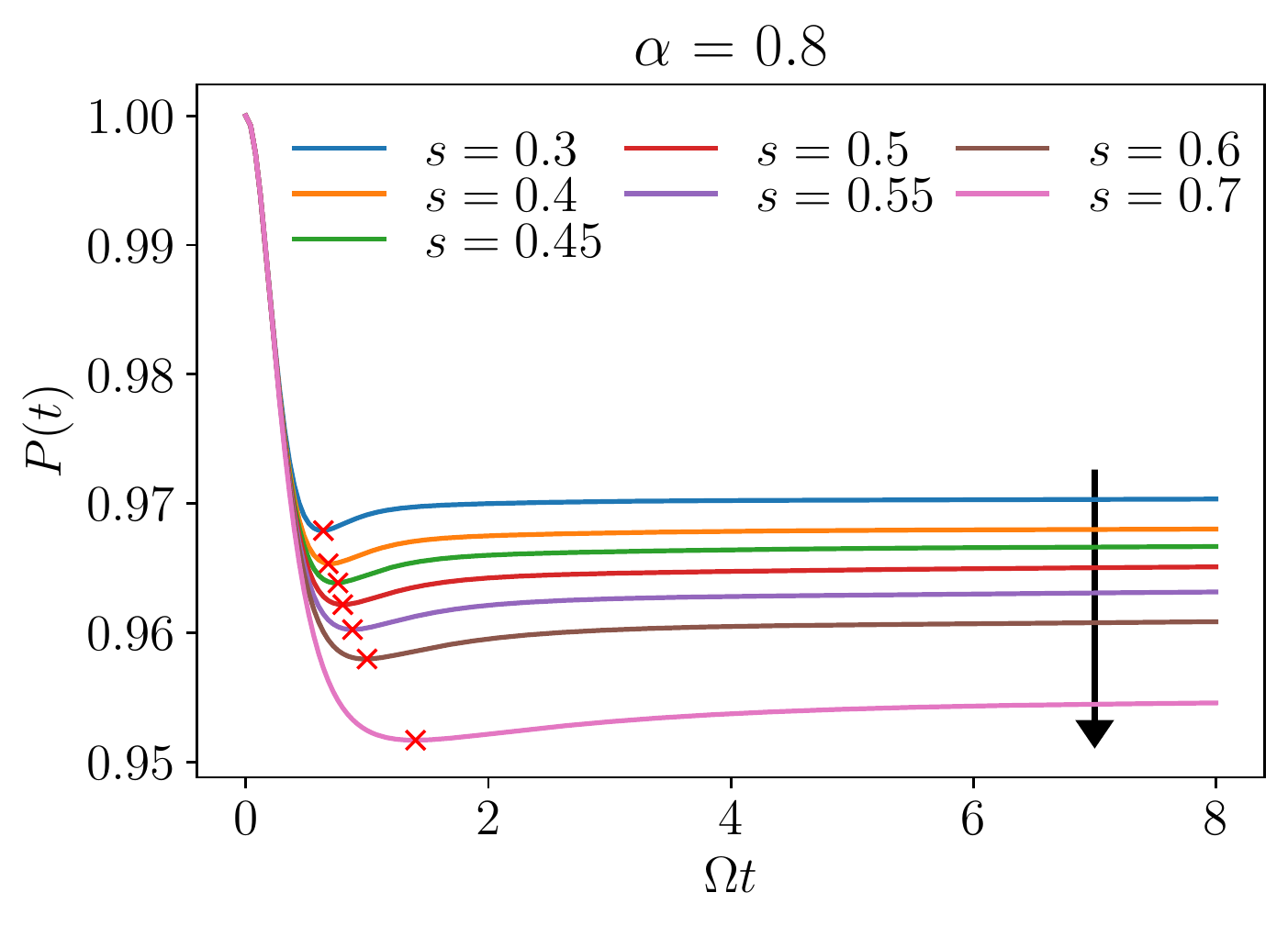}
\caption{Polarization $P(t)$ for $T=0$, $\alpha = 0.8$, and different spectral exponents $s$. The arrow intersects the lines in ascending order of spectral exponents. Local minima are marked with a red cross.}
\label{sectionplot_alpha_0.8}
\end{figure}

Next we decrease the spectral exponent $s$ for a strong but fixed system bath coupling strength $\alpha=0.8$ with dynamics in the pseudo-coherent phase from $s=0.7$ to $0.3$. Fig.~\ref{sectionplot_alpha_0.8} depicts the according time-dependent polarization $P(t)$. In all cases we observe pseudo-coherent dynamics with a single minimum (marked by a red cross). With decreasing $s$, the minimum shifts to earlier times. No qualitative change in the dynamic behaviour is found upon decreasing $s$ below $s=1/2$. For $s\lesssim 1/2$, Kast and Ankerhold \cite{kast_persistence_2013} have shown that no transition to incoherent dynamics occurs irrespective of the coupling strength.

\begin{figure}[h]
\includegraphics[width=86mm]{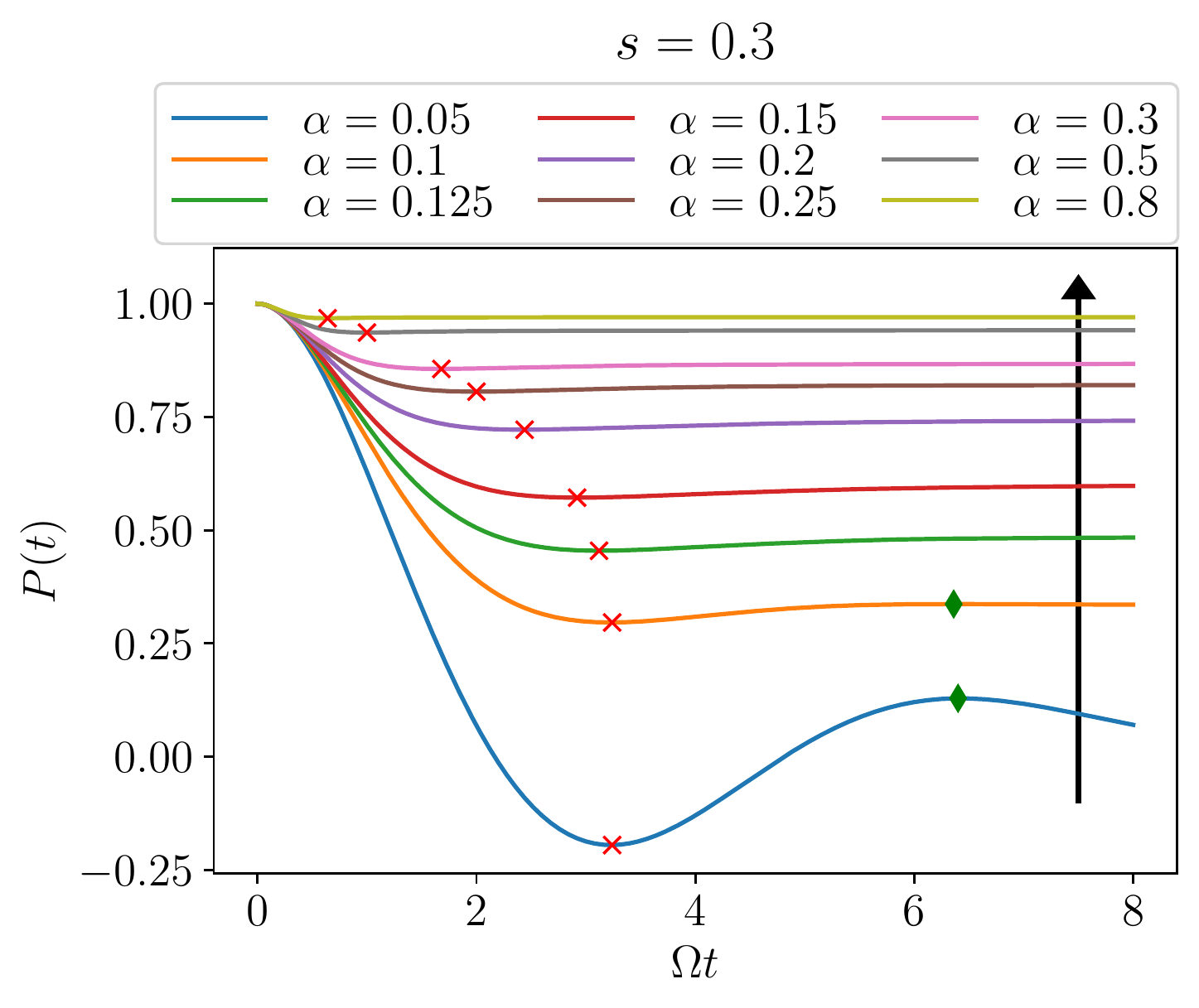}
\caption{Polarization $P(t)$ for $T=0$, $s=0.3$, and different coupling strengths $\alpha$. The arrow intersects the lines in ascending order of coupling strengths. Local minima (maxima) are marked by a red cross (green diamond).  }
\label{fig:sectionplot_s_0.3}
\end{figure}

Fig.~\ref{fig:sectionplot_s_0.3} shows the polarization for various coupling strengths for a fixed $s=0.3$. For couplings from $\alpha=0.8$ down to $\alpha=0.125$, we observe pseudo-coherent dynamics with only one minimum. For $\alpha = 0.1$ and $\alpha = 0.05$, we find coherent dynamics with a minimum and a maximum.

\begin{figure}[h] 
\includegraphics[width=86mm]{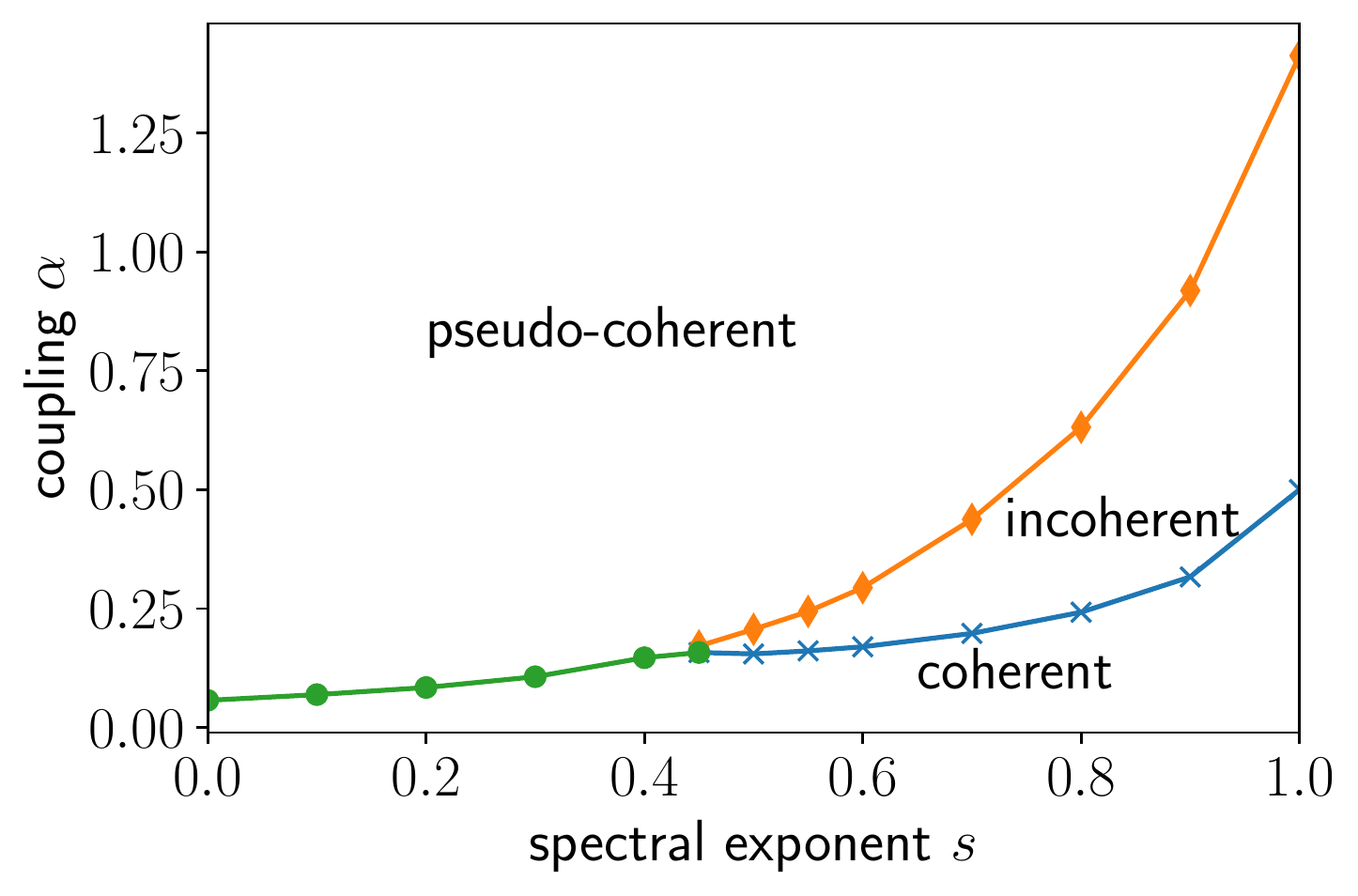}
\caption{Phase diagram of the (sub-)Ohmic spin-boson model at $T=0$ and $\omega_c  = 10 \Omega$. The symbols represent coupling strengths $\alpha (s)$ at which a transition occurs and are linearly interpolated for better visibility. Blue Cross: The first local minimum vanishes when increasing coupling out of the coherent to the incoherent domain. Orange diamond: A local minimum appears for $\Omega t< 8$ when increasing coupling strength out of the incoherent domain. Green circle: The first local maximum vanishes when going from the coherent to the pseudo-coherent domain.}
\label{fig:subohmic_phasediagram}
\end{figure}

\textit{Phase diagram}. --- %
Next we examine the full parameter space of $0\le s\le 1$ and $\alpha$. We observe sharp transitions between three dynamical phases: coherent dynamics, i.e. damped oscillatory behaviour with minima and maxima, incoherent dynamics, i.e. purely monotonic decay, and pseudo-coherent dynamics, i.e. a single minimum and subsequent decay into localization. The observed phase diagram is sketched in Fig.~\ref{fig:subohmic_phasediagram}.
For $\alpha = 0$, the time-dependent polarization is $P(t) = \cos (\Omega t )$ with its first local minimum at $\Omega t = \pi$. As the dynamics depends continuously on the parameters $\alpha$ and $s$, we can track this local minimum while increasing the coupling strength. For $s \gtrsim 0.45$, we find that this minimum vanishes if the coupling strength is sufficiently increased leading to the transition from coherent to incoherent dynamics at $\alpha_D(s)$ (depicted as blue crosses in Fig.~\ref{fig:subohmic_phasediagram}). Our results coincide with results shown in Fig.~5 of Ref.~\cite{duan_zero-temperature_2017} for $s>0.5$. 
Increasing the coupling strength further, we enter the pseudo-coherent phase at $\alpha_B(s)$ (depicted by orange diamonds). For $s\leq 0.45$ we track the first local maximum of the coherent phase, which vanishes in the transition to the pseudo-coherent regime (see Fig.~\ref{fig:sectionplot_s_0.3} for $s=0.3$). The green circles in Fig.~\ref{fig:subohmic_phasediagram} depict the coupling strength $\alpha_B(s)$ of the respective phase transition from coherent to pseudo-coherent dynamics. Note that while maxima in the dynamics of the pseudo-coherent phase at very long times, i.e. time-scales not determined by $\Omega$ nor $\omega_c$ (which we cannot strictly rule out due to finite simulation times), would add to the characteristics of the pseudo-coherent phase, they would not change the phase diagram which is determined by sharp transitions between different dynamic behaviour at times well within our simulation time window.

\begin{figure}[t]
\includegraphics[width=86mm]{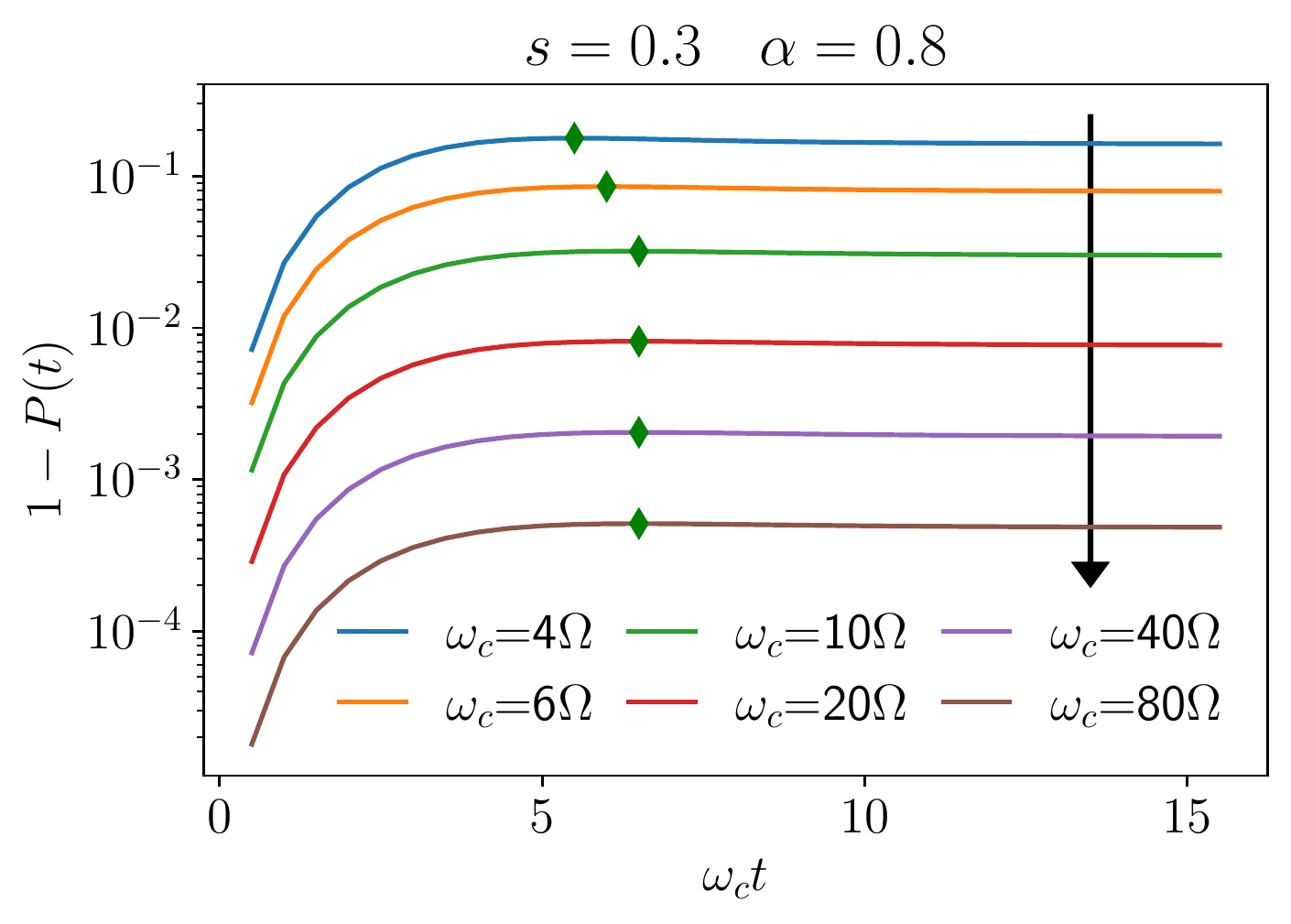}
\caption{Polarization $1 - P(t)$ for $T=0$, $\alpha = 0.8$, $s=0.3$, and different bath cut-off frequencies $\omega_c$. The arrow intersects the lines in ascending order of cut-off frequencies. Local maxima are marked with a green diamond. Note that the $\omega_c = 10 \Omega$ line corresponds to the $s=0.3$ line in Fig.~\ref{sectionplot_alpha_0.8}.}
\label{bath_dominated_oscillation}
\end{figure}

\textit{Pseudo-coherent phase}. --- %
Next, we investigate the pseudo-coherent phase in more detail. Fig.~\ref{bath_dominated_oscillation} shows the quantity $1-P(t)$ versus the rescaled time $\omega_c t$ for various $\omega_c$ for $s=0.3$ and $\alpha=0.8$. (Note that the minima in $P(t)$ are maxima in this plot). 
We observe that the maximum in $1 - P(t)$ (corresponding to a minimum in $P(t)$) for sufficiently large values of $\omega_c \gtrsim 10 \Omega$ occurs at a time $t_{\rm min}\simeq \text{const.}/\omega_c $. Hence, the observed oscillatory behaviour in this phase is purely bath-driven and therefore distinct from coherence, which generally refers to a behaviour inherent to the system. This furthermore suggests that for $\omega_c\rightarrow\infty$ the polarization $P(t)$ approaches 1 and the minimum is attained at $t_{\rm min}\rightarrow 0$. Thus, only incoherent fully localized behaviour emerges, i.e., the observed oscillatory behaviour in the pseudo-coherent phase is a result of the finite bath reaction time of $\mathcal{O} ( 1/\omega_c )$ only after which the system is drawn into localization. 
Similarly, purely dephasing sub-Ohmic reservoirs 
for small $s$ 
renormalize for increasing $\omega_c$ the oscillatory frequency always to larger values than its damping rate and the system frequency component in the total renormalized frequency vanishes in that regime \cite{nalbach_crossover_2013}. Thus, the dynamics  never becomes fully overdamped.

Because the oscillation frequency is renormalized in the case of a polarized bath that is initially equilibrated to a spin pointing up, the oscillatory behavior in Ref.~\cite{kast_persistence_2013} also shows maxima in the pseudo-coherent phase. Nevertheless, the dynamics in that case is  generated by the bath in the same way (for details, see Supplemental Material).

Increasing $\omega_c$ also changes the phase separation line $\alpha_B(s)$. To determine the phase separation line $\alpha_B(s)$, we simulate the polarization dynamics to roughly similar times, i.e., $\Omega t \simeq 10$ for every $\omega_c$. Thus, we are here limited to a fairly small regime of $\omega_c\le 50\Omega$ to ensure numerical accuracy.
For all studied values of $\omega_c$ we find that $\alpha_B(s)>\alpha_c(s)$ with the critical coupling strength $\alpha_c(s)$ for the thermodynamic $T=0$ phase transition to localization. We find that $\alpha_B(s)$ decreases with increasing $\omega_c$.

Our investigation allows to conclude that for $\omega_c\to\infty$ the pseudo-coherent phase turns fully incoherent. Thus, for $s\gtrsim 0.45$ it seems reasonable that $\alpha_B(s,\omega_c)\to\alpha_c(s,\omega_c)$ since for couplings smaller than $\alpha_B(s,\omega_c)$ the dynamics is already incoherent for all $\omega_c$. For $s\lesssim 0.45$ the situation is less clear and it remains open whether for $\omega_c\to\infty$ there is a coupling regime with damped coherent dynamics. We should point out that the limit of infinite $\omega_c$ is also not clear for $\alpha_c$, i.e. the thermodynamic localization transition, except for Ohmic bath spectra where $\alpha_c(s=1,\omega_c\to\infty)=1$. For $0\le s\le 1/2$ a variational approach yields $\alpha_c(s,\omega_c\to\infty) \to 0$ \cite{chin_generalized_2011}.

\begin{figure}[h]
\includegraphics[width=86mm]{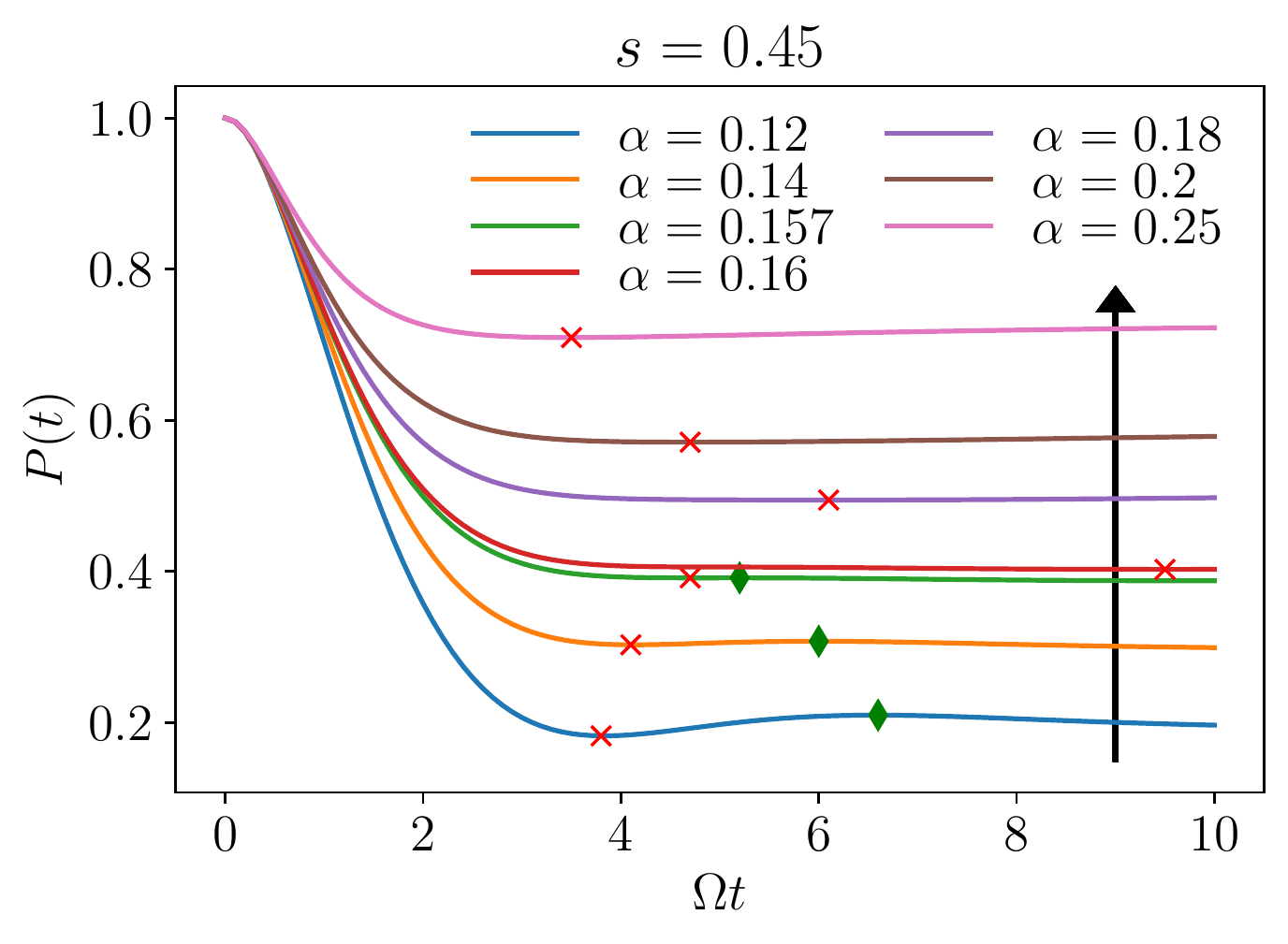}
\caption{Polarization $P(t)$ for $T=0$, $s=0.45$, and different coupling strengths $\alpha$. The arrow intersects the lines in ascending order of coupling strengths. Local minima (maxima) are marked with a red cross (green diamond). }
\label{fig:sectionplot_s_0.45}
\end{figure}

\textit{Spectral crossover}. --- %
Investigations of the thermodynamic localization phase transition \cite{winter_quantum_2009}, i.e. the according critical exponents, conclude that $s=1/2$ is the border between mean-field behaviour for $s<1/2$ and non mean-field for $s>1/2$. We find a crossover $s_{\rm cross}$ with a possible incoherent phase for $s>s_{\rm cross}$ and no incoherent phase for $s<s_{\rm cross}$ at $s_{\rm cross}\simeq 0.45$ (see Fig.~\ref{fig:subohmic_phasediagram}), which is consistent with the results of Duan et al. \cite{duan_zero-temperature_2017} that imply $s_{\rm cross} \lesssim 0.525$. 
Fig.~\ref{fig:sectionplot_s_0.45} shows the polarization for $T=0$, $s=0.45$, and different coupling strengths $\alpha$. For $\alpha=0.12$, we observe coherent dynamics with a minimum and maximum. For increasing couplings, i.e. $\alpha=0.14$ and $0.157$, both merge. For $\alpha=0.16$, a new single minimum at latest observed time appears which for further increasing coupling shifts to earlier times. 
For $s<0.45$ (see, for example, Fig.~\ref{fig:sectionplot_s_0.3}), the observed minimum in the pseudo-coherent phase emerges from the first minimum in the coherent phase. It remains an open question how this crossover $s$ depends on $\omega_c$ and in how far it is connected to the crossover from mean-field to non-mean-field type thermodynamics phase behaviour.

\textit{Conclusion}. --- %
By means of numerically exact real-time path integral simulations, we have studied the nonequilibrium dynamics of the (sub-)Ohmic spin-boson model. Analysing the polarization dynamics, we find for all spectral exponents $0\le s \le 1$ at strong coupling a pseudo-coherent phase whose hallmark is a single oscillatory minimum. In particular for $0.45\le s \le 1$, we observe this new dynamical phase at strong coupling beside the well-known coherent dynamics at weak coupling and incoherent dynamics at intermediate coupling. Oscillatory dynamics for $0\le s \le 1/2$ for all couplings has been observed by Kast and Ankerhold \cite{kast_persistence_2013} before. We show that for these $s$ there is nevertheless a transition from coherent (many minima and maxima in the polarization dynamics) to pseudo-coherent dynamics (with a single minimum). The frequency related to the oscillatory minimum in the pseudo-coherent phase is proportional to the bath cut-off frequency. Accordingly, this dynamics is not generated by the two-level system but by the reservoir and it turns incoherent for $\omega_c\rightarrow\infty$. We map the full dynamical phase diagram with now three distinct phases.  \\

F.O. is grateful for the kind hospitality at the Flatiron Institute where parts of this work have been carried out.



\bibliography{apssamp}

\providecommand{\noopsort}[1]{}\providecommand{\singleletter}[1]{#1}%
\begin{thebibliography}{25}%
\makeatletter
\providecommand \@ifxundefined [1]{%
 \@ifx{#1\undefined}
}%
\providecommand \@ifnum [1]{%
 \ifnum #1\expandafter \@firstoftwo
 \else \expandafter \@secondoftwo
 \fi
}%
\providecommand \@ifx [1]{%
 \ifx #1\expandafter \@firstoftwo
 \else \expandafter \@secondoftwo
 \fi
}%
\providecommand \natexlab [1]{#1}%
\providecommand \enquote  [1]{``#1''}%
\providecommand \bibnamefont  [1]{#1}%
\providecommand \bibfnamefont [1]{#1}%
\providecommand \citenamefont [1]{#1}%
\providecommand \href@noop [0]{\@secondoftwo}%
\providecommand \href [0]{\begingroup \@sanitize@url \@href}%
\providecommand \@href[1]{\@@startlink{#1}\@@href}%
\providecommand \@@href[1]{\endgroup#1\@@endlink}%
\providecommand \@sanitize@url [0]{\catcode `\\12\catcode `\$12\catcode
  `\&12\catcode `\#12\catcode `\^12\catcode `\_12\catcode `\%12\relax}%
\providecommand \@@startlink[1]{}%
\providecommand \@@endlink[0]{}%
\providecommand \url  [0]{\begingroup\@sanitize@url \@url }%
\providecommand \@url [1]{\endgroup\@href {#1}{\urlprefix }}%
\providecommand \urlprefix  [0]{URL }%
\providecommand \Eprint [0]{\href }%
\providecommand \doibase [0]{https://doi.org/}%
\providecommand \selectlanguage [0]{\@gobble}%
\providecommand \bibinfo  [0]{\@secondoftwo}%
\providecommand \bibfield  [0]{\@secondoftwo}%
\providecommand \translation [1]{[#1]}%
\providecommand \BibitemOpen [0]{}%
\providecommand \bibitemStop [0]{}%
\providecommand \bibitemNoStop [0]{.\EOS\space}%
\providecommand \EOS [0]{\spacefactor3000\relax}%
\providecommand \BibitemShut  [1]{\csname bibitem#1\endcsname}%
\let\auto@bib@innerbib\@empty
\bibitem [{\citenamefont {Leggett}\ \emph {et~al.}(1987)\citenamefont
  {Leggett}, \citenamefont {Chakravarty}, \citenamefont {Dorsey}, \citenamefont
  {Fisher}, \citenamefont {Garg},\ and\ \citenamefont
  {Zwerger}}]{leggett_dynamics_1987}%
  \BibitemOpen
  \bibfield  {author} {\bibinfo {author} {\bibfnamefont {A.~J.}\ \bibnamefont
  {Leggett}}, \bibinfo {author} {\bibfnamefont {S.}~\bibnamefont
  {Chakravarty}}, \bibinfo {author} {\bibfnamefont {A.~T.}\ \bibnamefont
  {Dorsey}}, \bibinfo {author} {\bibfnamefont {M.~P.~A.}\ \bibnamefont
  {Fisher}}, \bibinfo {author} {\bibfnamefont {A.}~\bibnamefont {Garg}},\ and\
  \bibinfo {author} {\bibfnamefont {W.}~\bibnamefont {Zwerger}},\ }\bibfield
  {title} {\bibinfo {title} {Dynamics of the dissipative two-state system},\
  }\href {https://doi.org/10.1103/RevModPhys.59.1} {\bibfield  {journal}
  {\bibinfo  {journal} {Rev. Mod. Phys.}\ }\textbf {\bibinfo {volume} {59}},\
  \bibinfo {pages} {1} (\bibinfo {year} {1987})}\BibitemShut {NoStop}%
\bibitem [{\citenamefont {Weiss}(2012)}]{weiss_quantum_2012}%
  \BibitemOpen
  \bibfield  {author} {\bibinfo {author} {\bibfnamefont {U.}~\bibnamefont
  {Weiss}},\ }\href {https://doi.org/10.1142/8334} {\emph {\bibinfo {title}
  {Quantum {Dissipative} {Systems}}}},\ \bibinfo {edition} {4th}\ ed.\
  (\bibinfo  {publisher} {World Scientific},\ \bibinfo {year}
  {2012})\BibitemShut {NoStop}%
\bibitem [{\citenamefont {Iles-Smith}\ \emph {et~al.}(2014)\citenamefont
  {Iles-Smith}, \citenamefont {Lambert},\ and\ \citenamefont
  {Nazir}}]{iles-smith_environmental_2014}%
  \BibitemOpen
  \bibfield  {author} {\bibinfo {author} {\bibfnamefont {J.}~\bibnamefont
  {Iles-Smith}}, \bibinfo {author} {\bibfnamefont {N.}~\bibnamefont
  {Lambert}},\ and\ \bibinfo {author} {\bibfnamefont {A.}~\bibnamefont
  {Nazir}},\ }\bibfield  {title} {\bibinfo {title} {Environmental dynamics,
  correlations, and the emergence of noncanonical equilibrium states in open
  quantum systems},\ }\href {https://doi.org/10.1103/PhysRevA.90.032114}
  {\bibfield  {journal} {\bibinfo  {journal} {Phys. Rev. A}\ }\textbf {\bibinfo
  {volume} {90}},\ \bibinfo {pages} {032114} (\bibinfo {year}
  {2014})}\BibitemShut {NoStop}%
\bibitem [{\citenamefont {Maguire}\ \emph {et~al.}(2019)\citenamefont
  {Maguire}, \citenamefont {Iles-Smith},\ and\ \citenamefont
  {Nazir}}]{maguire_environmental_2019}%
  \BibitemOpen
  \bibfield  {author} {\bibinfo {author} {\bibfnamefont {H.}~\bibnamefont
  {Maguire}}, \bibinfo {author} {\bibfnamefont {J.}~\bibnamefont
  {Iles-Smith}},\ and\ \bibinfo {author} {\bibfnamefont {A.}~\bibnamefont
  {Nazir}},\ }\bibfield  {title} {\bibinfo {title} {Environmental
  {Nonadditivity} and {Franck}-{Condon} physics in {Nonequilibrium} {Quantum}
  {Systems}},\ }\href {https://doi.org/10.1103/PhysRevLett.123.093601}
  {\bibfield  {journal} {\bibinfo  {journal} {Phys. Rev. Lett.}\ }\textbf
  {\bibinfo {volume} {123}},\ \bibinfo {pages} {093601} (\bibinfo {year}
  {2019})}\BibitemShut {NoStop}%
\bibitem [{\citenamefont {Anders}\ \emph {et~al.}(2007)\citenamefont {Anders},
  \citenamefont {Bulla},\ and\ \citenamefont
  {Vojta}}]{anders_equilibrium_2007}%
  \BibitemOpen
  \bibfield  {author} {\bibinfo {author} {\bibfnamefont {F.~B.}\ \bibnamefont
  {Anders}}, \bibinfo {author} {\bibfnamefont {R.}~\bibnamefont {Bulla}},\ and\
  \bibinfo {author} {\bibfnamefont {M.}~\bibnamefont {Vojta}},\ }\bibfield
  {title} {\bibinfo {title} {Equilibrium and {Nonequilibrium} {Dynamics} of the
  {Sub}-{Ohmic} {Spin}-{Boson} {Model}},\ }\href
  {https://doi.org/10.1103/PhysRevLett.98.210402} {\bibfield  {journal}
  {\bibinfo  {journal} {Phys. Rev. Lett.}\ }\textbf {\bibinfo {volume} {98}},\
  \bibinfo {pages} {210402} (\bibinfo {year} {2007})}\BibitemShut {NoStop}%
\bibitem [{\citenamefont {Winter}\ \emph {et~al.}(2009)\citenamefont {Winter},
  \citenamefont {Rieger}, \citenamefont {Vojta},\ and\ \citenamefont
  {Bulla}}]{winter_quantum_2009}%
  \BibitemOpen
  \bibfield  {author} {\bibinfo {author} {\bibfnamefont {A.}~\bibnamefont
  {Winter}}, \bibinfo {author} {\bibfnamefont {H.}~\bibnamefont {Rieger}},
  \bibinfo {author} {\bibfnamefont {M.}~\bibnamefont {Vojta}},\ and\ \bibinfo
  {author} {\bibfnamefont {R.}~\bibnamefont {Bulla}},\ }\bibfield  {title}
  {\bibinfo {title} {Quantum {Phase} {Transition} in the {Sub}-{Ohmic}
  {Spin}-{Boson} {Model}: {Quantum} {Monte} {Carlo} {Study} with a {Continuous}
  {Imaginary} {Time} {Cluster} {Algorithm}},\ }\href
  {https://doi.org/10.1103/PhysRevLett.102.030601} {\bibfield  {journal}
  {\bibinfo  {journal} {Phys. Rev. Lett.}\ }\textbf {\bibinfo {volume} {102}},\
  \bibinfo {pages} {030601} (\bibinfo {year} {2009})}\BibitemShut {NoStop}%
\bibitem [{\citenamefont {Alvermann}\ and\ \citenamefont
  {Fehske}(2009)}]{alvermann_sparse_2009}%
  \BibitemOpen
  \bibfield  {author} {\bibinfo {author} {\bibfnamefont {A.}~\bibnamefont
  {Alvermann}}\ and\ \bibinfo {author} {\bibfnamefont {H.}~\bibnamefont
  {Fehske}},\ }\bibfield  {title} {\bibinfo {title} {Sparse {Polynomial}
  {Space} {Approach} to {Dissipative} {Quantum} {Systems}: {Application} to the
  {Sub}-{Ohmic} {Spin}-{Boson} {Model}},\ }\href
  {https://doi.org/10.1103/PhysRevLett.102.150601} {\bibfield  {journal}
  {\bibinfo  {journal} {Phys. Rev. Lett.}\ }\textbf {\bibinfo {volume} {102}},\
  \bibinfo {pages} {150601} (\bibinfo {year} {2009})}\BibitemShut {NoStop}%
\bibitem [{\citenamefont {Si}\ \emph {et~al.}(2001)\citenamefont {Si},
  \citenamefont {Rabello}, \citenamefont {Ingersent},\ and\ \citenamefont
  {Smith}}]{si_locally_2001}%
  \BibitemOpen
  \bibfield  {author} {\bibinfo {author} {\bibfnamefont {Q.}~\bibnamefont
  {Si}}, \bibinfo {author} {\bibfnamefont {S.}~\bibnamefont {Rabello}},
  \bibinfo {author} {\bibfnamefont {K.}~\bibnamefont {Ingersent}},\ and\
  \bibinfo {author} {\bibfnamefont {J.~L.}\ \bibnamefont {Smith}},\ }\bibfield
  {title} {\bibinfo {title} {Locally critical quantum phase transitions in
  strongly correlated metals},\ }\href {https://doi.org/10.1038/35101507}
  {\bibfield  {journal} {\bibinfo  {journal} {Nature}\ }\textbf {\bibinfo
  {volume} {413}},\ \bibinfo {pages} {804} (\bibinfo {year}
  {2001})}\BibitemShut {NoStop}%
\bibitem [{\citenamefont {Gegenwart}\ \emph {et~al.}(2007)\citenamefont
  {Gegenwart}, \citenamefont {Westerkamp}, \citenamefont {Krellner},
  \citenamefont {Tokiwa}, \citenamefont {Paschen}, \citenamefont {Geibel},
  \citenamefont {Steglich}, \citenamefont {Abrahams},\ and\ \citenamefont
  {Si}}]{gegenwart_multiple_2007}%
  \BibitemOpen
  \bibfield  {author} {\bibinfo {author} {\bibfnamefont {P.}~\bibnamefont
  {Gegenwart}}, \bibinfo {author} {\bibfnamefont {T.}~\bibnamefont
  {Westerkamp}}, \bibinfo {author} {\bibfnamefont {C.}~\bibnamefont
  {Krellner}}, \bibinfo {author} {\bibfnamefont {Y.}~\bibnamefont {Tokiwa}},
  \bibinfo {author} {\bibfnamefont {S.}~\bibnamefont {Paschen}}, \bibinfo
  {author} {\bibfnamefont {C.}~\bibnamefont {Geibel}}, \bibinfo {author}
  {\bibfnamefont {F.}~\bibnamefont {Steglich}}, \bibinfo {author}
  {\bibfnamefont {E.}~\bibnamefont {Abrahams}},\ and\ \bibinfo {author}
  {\bibfnamefont {Q.}~\bibnamefont {Si}},\ }\bibfield  {title} {\bibinfo
  {title} {Multiple energy scales at a quantum critical point},\ }\href
  {https://doi.org/10.1126/science.1136020} {\bibfield  {journal} {\bibinfo
  {journal} {Science}\ }\textbf {\bibinfo {volume} {315}},\ \bibinfo {pages}
  {969} (\bibinfo {year} {2007})}\BibitemShut {NoStop}%
\bibitem [{\citenamefont {Gröblacher}\ \emph {et~al.}(2015)\citenamefont
  {Gröblacher}, \citenamefont {Trubarov}, \citenamefont {Prigge},
  \citenamefont {Cole}, \citenamefont {Aspelmeyer},\ and\ \citenamefont
  {Eisert}}]{groblacher_observation_2015}%
  \BibitemOpen
  \bibfield  {author} {\bibinfo {author} {\bibfnamefont {S.}~\bibnamefont
  {Gröblacher}}, \bibinfo {author} {\bibfnamefont {A.}~\bibnamefont
  {Trubarov}}, \bibinfo {author} {\bibfnamefont {N.}~\bibnamefont {Prigge}},
  \bibinfo {author} {\bibfnamefont {G.~D.}\ \bibnamefont {Cole}}, \bibinfo
  {author} {\bibfnamefont {M.}~\bibnamefont {Aspelmeyer}},\ and\ \bibinfo
  {author} {\bibfnamefont {J.}~\bibnamefont {Eisert}},\ }\bibfield  {title}
  {\bibinfo {title} {Observation of non-{Markovian} micromechanical {Brownian}
  motion},\ }\href {https://doi.org/10.1038/ncomms8606} {\bibfield  {journal}
  {\bibinfo  {journal} {Nat Commun}\ }\textbf {\bibinfo {volume} {6}},\
  \bibinfo {pages} {7606} (\bibinfo {year} {2015})}\BibitemShut {NoStop}%
\bibitem [{\citenamefont {Astafiev}\ \emph {et~al.}(2004)\citenamefont
  {Astafiev}, \citenamefont {Pashkin}, \citenamefont {Nakamura}, \citenamefont
  {Yamamoto},\ and\ \citenamefont {Tsai}}]{AstafievPRL2004}%
  \BibitemOpen
  \bibfield  {author} {\bibinfo {author} {\bibfnamefont {O.}~\bibnamefont
  {Astafiev}}, \bibinfo {author} {\bibfnamefont {Y.~A.}\ \bibnamefont
  {Pashkin}}, \bibinfo {author} {\bibfnamefont {Y.}~\bibnamefont {Nakamura}},
  \bibinfo {author} {\bibfnamefont {T.}~\bibnamefont {Yamamoto}},\ and\
  \bibinfo {author} {\bibfnamefont {J.~S.}\ \bibnamefont {Tsai}},\ }\bibfield
  {title} {\bibinfo {title} {Quantum noise in the josephson charge qubit},\
  }\href {https://doi.org/10.1103/PhysRevLett.93.267007} {\bibfield  {journal}
  {\bibinfo  {journal} {Phys. Rev. Lett.}\ }\textbf {\bibinfo {volume} {93}},\
  \bibinfo {pages} {267007} (\bibinfo {year} {2004})}\BibitemShut {NoStop}%
\bibitem [{\citenamefont {Shnirman}\ \emph {et~al.}(2005)\citenamefont
  {Shnirman}, \citenamefont {Sch\"on}, \citenamefont {Martin},\ and\
  \citenamefont {Makhlin}}]{ShnirmanPRL2005}%
  \BibitemOpen
  \bibfield  {author} {\bibinfo {author} {\bibfnamefont {A.}~\bibnamefont
  {Shnirman}}, \bibinfo {author} {\bibfnamefont {G.}~\bibnamefont {Sch\"on}},
  \bibinfo {author} {\bibfnamefont {I.}~\bibnamefont {Martin}},\ and\ \bibinfo
  {author} {\bibfnamefont {Y.}~\bibnamefont {Makhlin}},\ }\bibfield  {title}
  {\bibinfo {title} {Low- and high-frequency noise from coherent two-level
  systems},\ }\href {https://doi.org/10.1103/PhysRevLett.94.127002} {\bibfield
  {journal} {\bibinfo  {journal} {Phys. Rev. Lett.}\ }\textbf {\bibinfo
  {volume} {94}},\ \bibinfo {pages} {127002} (\bibinfo {year}
  {2005})}\BibitemShut {NoStop}%
\bibitem [{\citenamefont {You}\ \emph {et~al.}(2021)\citenamefont {You},
  \citenamefont {Clerk},\ and\ \citenamefont {Koch}}]{YouPRR2021}%
  \BibitemOpen
  \bibfield  {author} {\bibinfo {author} {\bibfnamefont {X.}~\bibnamefont
  {You}}, \bibinfo {author} {\bibfnamefont {A.~A.}\ \bibnamefont {Clerk}},\
  and\ \bibinfo {author} {\bibfnamefont {J.}~\bibnamefont {Koch}},\ }\bibfield
  {title} {\bibinfo {title} {Positive- and negative-frequency noise from an
  ensemble of two-level fluctuators},\ }\href
  {https://doi.org/10.1103/PhysRevResearch.3.013045} {\bibfield  {journal}
  {\bibinfo  {journal} {Phys. Rev. Research}\ }\textbf {\bibinfo {volume}
  {3}},\ \bibinfo {pages} {013045} (\bibinfo {year} {2021})}\BibitemShut
  {NoStop}%
\bibitem [{\citenamefont {Chin}\ and\ \citenamefont
  {Turlakov}(2006)}]{chin_coherent-incoherent_2006}%
  \BibitemOpen
  \bibfield  {author} {\bibinfo {author} {\bibfnamefont {A.}~\bibnamefont
  {Chin}}\ and\ \bibinfo {author} {\bibfnamefont {M.}~\bibnamefont
  {Turlakov}},\ }\bibfield  {title} {\bibinfo {title} {Coherent-incoherent
  transition in the sub-{Ohmic} spin-boson model},\ }\href
  {https://doi.org/10.1103/PhysRevB.73.075311} {\bibfield  {journal} {\bibinfo
  {journal} {Phys. Rev. B}\ }\textbf {\bibinfo {volume} {73}},\ \bibinfo
  {pages} {075311} (\bibinfo {year} {2006})}\BibitemShut {NoStop}%
\bibitem [{\citenamefont {Nalbach}\ and\ \citenamefont
  {Thorwart}(2010)}]{nalbach_ultraslow_2010}%
  \BibitemOpen
  \bibfield  {author} {\bibinfo {author} {\bibfnamefont {P.}~\bibnamefont
  {Nalbach}}\ and\ \bibinfo {author} {\bibfnamefont {M.}~\bibnamefont
  {Thorwart}},\ }\bibfield  {title} {\bibinfo {title} {Ultraslow quantum
  dynamics in a sub-{Ohmic} heat bath},\ }\href
  {https://doi.org/10.1103/PhysRevB.81.054308} {\bibfield  {journal} {\bibinfo
  {journal} {Phys. Rev. B}\ }\textbf {\bibinfo {volume} {81}},\ \bibinfo
  {pages} {054308} (\bibinfo {year} {2010})}\BibitemShut {NoStop}%
\bibitem [{\citenamefont {Kast}\ and\ \citenamefont
  {Ankerhold}(2013)}]{kast_persistence_2013}%
  \BibitemOpen
  \bibfield  {author} {\bibinfo {author} {\bibfnamefont {D.}~\bibnamefont
  {Kast}}\ and\ \bibinfo {author} {\bibfnamefont {J.}~\bibnamefont
  {Ankerhold}},\ }\bibfield  {title} {\bibinfo {title} {Persistence of
  {Coherent} {Quantum} {Dynamics} at {Strong} {Dissipation}},\ }\href
  {https://doi.org/10.1103/PhysRevLett.110.010402} {\bibfield  {journal}
  {\bibinfo  {journal} {Phys. Rev. Lett.}\ }\textbf {\bibinfo {volume} {110}},\
  \bibinfo {pages} {010402} (\bibinfo {year} {2013})}\BibitemShut {NoStop}%
\bibitem [{\citenamefont {Duan}\ \emph {et~al.}(2017)\citenamefont {Duan},
  \citenamefont {Tang}, \citenamefont {Cao},\ and\ \citenamefont
  {Wu}}]{duan_zero-temperature_2017}%
  \BibitemOpen
  \bibfield  {author} {\bibinfo {author} {\bibfnamefont {C.}~\bibnamefont
  {Duan}}, \bibinfo {author} {\bibfnamefont {Z.}~\bibnamefont {Tang}}, \bibinfo
  {author} {\bibfnamefont {J.}~\bibnamefont {Cao}},\ and\ \bibinfo {author}
  {\bibfnamefont {J.}~\bibnamefont {Wu}},\ }\bibfield  {title} {\bibinfo
  {title} {Zero-temperature localization in a sub-{Ohmic} spin-boson model
  investigated by an extended hierarchy equation of motion},\ }\href
  {https://doi.org/10.1103/PhysRevB.95.214308} {\bibfield  {journal} {\bibinfo
  {journal} {Phys. Rev. B}\ }\textbf {\bibinfo {volume} {95}},\ \bibinfo
  {pages} {214308} (\bibinfo {year} {2017})}\BibitemShut {NoStop}%
\bibitem [{\citenamefont {Nalbach}\ and\ \citenamefont
  {Thorwart}(2013)}]{nalbach_crossover_2013}%
  \BibitemOpen
  \bibfield  {author} {\bibinfo {author} {\bibfnamefont {P.}~\bibnamefont
  {Nalbach}}\ and\ \bibinfo {author} {\bibfnamefont {M.}~\bibnamefont
  {Thorwart}},\ }\bibfield  {title} {\bibinfo {title} {Crossover from coherent
  to incoherent quantum dynamics due to sub-{Ohmic} dephasing},\ }\href
  {https://doi.org/10.1103/PhysRevB.87.014116} {\bibfield  {journal} {\bibinfo
  {journal} {Phys. Rev. B}\ }\textbf {\bibinfo {volume} {87}},\ \bibinfo
  {pages} {014116} (\bibinfo {year} {2013})}\BibitemShut {NoStop}%
\bibitem [{\citenamefont {Makri}(1995)}]{makri_numerical_1995}%
  \BibitemOpen
  \bibfield  {author} {\bibinfo {author} {\bibfnamefont {N.}~\bibnamefont
  {Makri}},\ }\bibfield  {title} {\bibinfo {title} {Numerical path integral
  techniques for long time dynamics of quantum dissipative systems},\ }\href
  {https://doi.org/10.1063/1.531046} {\bibfield  {journal} {\bibinfo  {journal}
  {J. Math. Phys.}\ }\textbf {\bibinfo {volume} {36}},\ \bibinfo {pages} {2430}
  (\bibinfo {year} {1995})}\BibitemShut {NoStop}%
\bibitem [{\citenamefont {Makri}\ and\ \citenamefont
  {Makarov}(1995{\natexlab{a}})}]{makri_tensor_1995}%
  \BibitemOpen
  \bibfield  {author} {\bibinfo {author} {\bibfnamefont {N.}~\bibnamefont
  {Makri}}\ and\ \bibinfo {author} {\bibfnamefont {D.~E.}\ \bibnamefont
  {Makarov}},\ }\bibfield  {title} {\bibinfo {title} {Tensor propagator for
  iterative quantum time evolution of reduced density matrices. {I}.
  {Theory}},\ }\href {https://doi.org/10.1063/1.469508} {\bibfield  {journal}
  {\bibinfo  {journal} {The Journal of Chemical Physics}\ }\textbf {\bibinfo
  {volume} {102}},\ \bibinfo {pages} {4600} (\bibinfo {year}
  {1995}{\natexlab{a}})}\BibitemShut {NoStop}%
\bibitem [{\citenamefont {Makri}\ and\ \citenamefont
  {Makarov}(1995{\natexlab{b}})}]{makri_tensor_1995-1}%
  \BibitemOpen
  \bibfield  {author} {\bibinfo {author} {\bibfnamefont {N.}~\bibnamefont
  {Makri}}\ and\ \bibinfo {author} {\bibfnamefont {D.~E.}\ \bibnamefont
  {Makarov}},\ }\bibfield  {title} {\bibinfo {title} {Tensor propagator for
  iterative quantum time evolution of reduced density matrices. {II}.
  {Numerical} methodology},\ }\href {https://doi.org/10.1063/1.469509}
  {\bibfield  {journal} {\bibinfo  {journal} {The Journal of Chemical Physics}\
  }\textbf {\bibinfo {volume} {102}},\ \bibinfo {pages} {4611} (\bibinfo {year}
  {1995}{\natexlab{b}})}\BibitemShut {NoStop}%
\bibitem [{\citenamefont {Makri}(2020)}]{makri_small_2020-1}%
  \BibitemOpen
  \bibfield  {author} {\bibinfo {author} {\bibfnamefont {N.}~\bibnamefont
  {Makri}},\ }\bibfield  {title} {\bibinfo {title} {Small {Matrix} {Path}
  {Integral} for {System}-{Bath} {Dynamics}},\ }\href
  {https://doi.org/10.1021/acs.jctc.0c00039} {\bibfield  {journal} {\bibinfo
  {journal} {J. Chem. Theory Comput.}\ }\textbf {\bibinfo {volume} {16}},\
  \bibinfo {pages} {4038} (\bibinfo {year} {2020})}\BibitemShut {NoStop}%
\bibitem [{\citenamefont {Strathearn}\ \emph {et~al.}(2018)\citenamefont
  {Strathearn}, \citenamefont {Kirton}, \citenamefont {Kilda}, \citenamefont
  {Keeling},\ and\ \citenamefont {Lovett}}]{strathearn_efficient_2018}%
  \BibitemOpen
  \bibfield  {author} {\bibinfo {author} {\bibfnamefont {A.}~\bibnamefont
  {Strathearn}}, \bibinfo {author} {\bibfnamefont {P.}~\bibnamefont {Kirton}},
  \bibinfo {author} {\bibfnamefont {D.}~\bibnamefont {Kilda}}, \bibinfo
  {author} {\bibfnamefont {J.}~\bibnamefont {Keeling}},\ and\ \bibinfo {author}
  {\bibfnamefont {B.~W.}\ \bibnamefont {Lovett}},\ }\bibfield  {title}
  {\bibinfo {title} {Efficient non-{Markovian} quantum dynamics using
  time-evolving matrix product operators},\ }\href
  {https://doi.org/10.1038/s41467-018-05617-3} {\bibfield  {journal} {\bibinfo
  {journal} {Nat Commun}\ }\textbf {\bibinfo {volume} {9}},\ \bibinfo {pages}
  {3322} (\bibinfo {year} {2018})}\BibitemShut {NoStop}%
\bibitem [{\citenamefont {Strathearn}(2020)}]{strathearn_modelling_2020}%
  \BibitemOpen
  \bibfield  {author} {\bibinfo {author} {\bibfnamefont {A.}~\bibnamefont
  {Strathearn}},\ }\href {https://doi.org/10.1007/978-3-030-54975-6} {\emph
  {\bibinfo {title} {Modelling {Non}-{Markovian} {Quantum} {Systems} {Using}
  {Tensor} {Networks}}}},\ Springer {Theses}\ (\bibinfo  {publisher} {Springer
  International Publishing},\ \bibinfo {address} {Cham},\ \bibinfo {year}
  {2020})\BibitemShut {NoStop}%
\bibitem [{\citenamefont {Chin}\ \emph {et~al.}(2011)\citenamefont {Chin},
  \citenamefont {Prior}, \citenamefont {Huelga},\ and\ \citenamefont
  {Plenio}}]{chin_generalized_2011}%
  \BibitemOpen
  \bibfield  {author} {\bibinfo {author} {\bibfnamefont {A.~W.}\ \bibnamefont
  {Chin}}, \bibinfo {author} {\bibfnamefont {J.}~\bibnamefont {Prior}},
  \bibinfo {author} {\bibfnamefont {S.~F.}\ \bibnamefont {Huelga}},\ and\
  \bibinfo {author} {\bibfnamefont {M.~B.}\ \bibnamefont {Plenio}},\ }\bibfield
   {title} {\bibinfo {title} {Generalized {Polaron} {Ansatz} for the {Ground}
  {State} of the {Sub}-{Ohmic} {Spin}-{Boson} {Model}: {An} {Analytic} {Theory}
  of the {Localization} {Transition}},\ }\href
  {https://doi.org/10.1103/PhysRevLett.107.160601} {\bibfield  {journal}
  {\bibinfo  {journal} {Phys. Rev. Lett.}\ }\textbf {\bibinfo {volume} {107}},\
  \bibinfo {pages} {160601} (\bibinfo {year} {2011})}\BibitemShut {NoStop}%
\end{thebibliography}%

\end{document}


\preprint{APS}

\title{Supplemental Material to\\ The Hidden Phase of the Spin-Boson Model}

\author{Florian Otterpohl $^{1,2}$}
\author{Peter Nalbach$^3$}
\author{Michael Thorwart$^2$}
\affiliation{$^1$Center for Computational Quantum Physics, Flatiron Institute, New York, New York 10010,
USA \\ $^2$I. Institut f\"ur Theoretische Physik, Universit\"at Hamburg, Notkestr.\ 9, 22607 Hamburg, Germany \\ $3$Fachbereich Wirtschaft \& Informationstechnik, Westfälische Hochschule, Münsterstr.\ 265, 46397 Bocholt,  Germany}

\date{\today}

\begin{abstract}
In this Supplemental Material, we provide a summary of the time-evolving matrix product operator (TEMPO) method as used in the main text and demonstrate the numerical convergence with respect to long simulation times. In addition, we also consider  initial conditions of a polarized bath and show that the dynamics is qualitatively very similar to the unpolarized initial conditions used in the main text.
\end{abstract}

\maketitle

\setcounter{equation}{0}
\setcounter{figure}{0}
\setcounter{table}{0}
\setcounter{page}{1}
\makeatletter
\renewcommand{\theequation}{S\arabic{equation}}
\renewcommand{\thefigure}{S\arabic{figure}}
\renewcommand{\thetable}{S\arabic{table}}
\renewcommand{\bibnumfmt}[1]{[S#1]}
\renewcommand{\citenumfont}[1]{S#1}

%
%
%
\section{Introduction}
In the main text, we compute the dynamics of the spin-boson model by use of the quasi-adiabatic propagator path integral (QUAPI) [19-24], which expresses the time evolution of the reduced density matrix in terms of a discrete path sum 
\begin{eqnarray}
\label{red}
\tilde{\rho}_{s_N^{\pm}} (t) && \equiv{} \operatorname{tr}_{\text{bath}} \Braket{ s_N^+ |e^{-i \hat{H} t} \tilde{\rho} (0) \rho_{B}(0) e^{i \hat{H} t} | s_N^- } \nonumber \\
&&=\sum_{s_{0}^\pm , ... , s_{N-1}^\pm \in \lbrace \uparrow , \downarrow \rbrace }  \prod_{j=1}^{N} \left( G_{s_j^\pm , s_{j-1}^\pm} \right)  \prod_{k = 0}^{N} \prod_{k^{\prime}=0}^{k} \left( F_{s_k^\pm , s_{k^\prime}^\pm}^{(k k^{\prime})} \right) \tilde{\rho}_{s_0^{\pm}} (0) + \mathcal{O} (\Delta t^2) \, ,
\end{eqnarray}
where $t = N \Delta t$. The system propagator is given by
\begin{equation}
    G_{s_{j}^\pm,s_{j-1}^\pm} = \Braket{s_{j}^+|e^{-i\hat{H}_S \Delta t}|s_{j-1}^+} \Braket{s_{j-1}^-|e^{i\hat{H}_S \Delta t}|s_{j}^-}\, ,
\end{equation}
and the bath influence functionals by
\begin{equation}
    F_{s_{k}^\pm s_{k^{\prime }}^\pm}^{(k k^{\prime })} = \exp \left[ - (s_k^+ - s_k^- )\left( \eta_{kk^{\prime}}  s_{k^\prime}^+ - \eta^*_{kk^{\prime}}  s_{k^\prime}^- \right) \right]\, ,
\end{equation}
where the coefficients $\eta_{k k^\prime}$ are functionals of the bath spectral density and are given in Ref.~[20]. Note that Eq.~(\ref{red}) is exact in the limit $\Delta t \rightarrow 0$.

\section{Time-Evolving Matrix Product Operator}
\label{tempo}
\begin{figure}[t!]
\centering
\begin{tabular}{cc}
\includegraphics[width=0.36\textwidth]{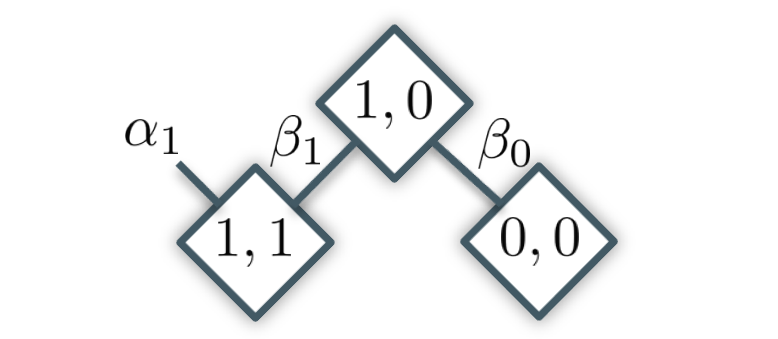}&
\includegraphics[width=0.58\textwidth]{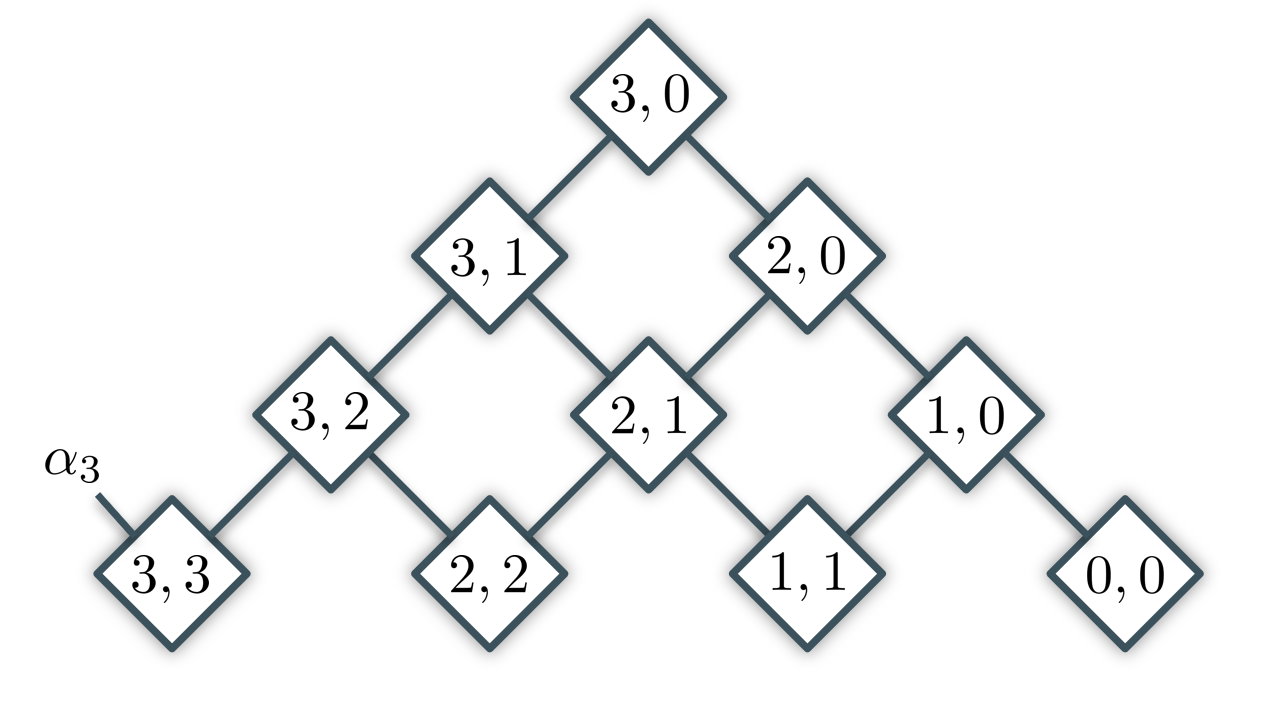} \\
\textbf{(a)}   & \textbf{(b)}     \\
\end{tabular}
\caption{Graphical representations of (a) Eq.~(\ref{path_I_n1}). Left, top, and right squares represent $\delta_{\alpha_1 \beta_1} I_{\alpha_1 \alpha_1}$, $I_{\beta_1 \beta_0}$, and $\sum_{\alpha_0} \delta_{\alpha_0 \beta_0} I_{\alpha_0 \alpha_0} \tilde{\rho}_{\alpha_0} $,  respectively. (b) Eq.~(\ref{path_I}) for $N=3$. Squares labeled with $k_1 , k_2$ consist of $I_{\alpha_{k_1} \alpha_{k_2}}$ and up to two Kronecker deltas that are of the form $\delta_{\alpha_{k_1} \beta_{k_1}}$ or $\delta_{\alpha_{k_1} \beta_{k_2}}$. The only exception of this is the bottom right square, which is given by $\sum_{\alpha_0} \delta_{\alpha_0 \beta_0} I_{\alpha_0 \alpha_0} \tilde{\rho}_{\alpha_0} $.}
\label{fig:tensordiag}
\end{figure}

As shown by Strathearn et al.~[23,24], the path sum in Eq.~(\ref{red}) can be rewritten in terms of a tensor network. By using the short-hand notation $s^\pm \equiv  \alpha\in \lbrace \upuparrows ,  \downdownarrows , \updownarrows , \downuparrows \rbrace$ and 
\begin{align}
I_{\alpha_{k_1} \alpha_{k_2}} :=
\begin{cases*}
          G_{\alpha_{k_1} \alpha_{k_2}}  F_{\alpha_{k_1} \alpha_{k_2}}^{(k_1 k_2)}
 & if   $\, k_1 - k_2  = 1$ \,  \\
                     F_{\alpha_{k_1} \alpha_{k_2}}^{(k_1 k_2)}
 & else, $ $
                 \end{cases*}
\end{align}
Eq.~(\ref{red}) becomes
\begin{align}
\label{path_I}
\tilde{\rho}_{\alpha_N} (t)= \sum_{\alpha_0 ,...,\alpha_{N-1}}  \prod_{k=0}^N \prod_{k^{\prime}=0}^{k} \left( I_{\alpha_{k} \alpha_{k^{\prime}}} \right) \tilde{\rho}_{\alpha_0} (0) \, .
\end{align}
Next, we consider the case $N=1$ of Eq.~(\ref{path_I}), which we rewrite using Kronecker deltas such that it can be drawn as a tensor diagram (see Fig.~\ref{fig:tensordiag} (a)):
\begin{eqnarray}
\label{path_I_n1}
\sum_{\alpha_0} I_{\alpha_1 \alpha_1} I_{\alpha_1 \alpha_0}  I_{\alpha_0 \alpha_0} \tilde{\rho}_{\alpha_0}  &&=  \sum_{\alpha_0} I_{\alpha_1 \alpha_1} \sum_{\beta_0} I_{\alpha_1 \beta_0} \delta_{\alpha_0 \beta_0} I_{\alpha_0 \alpha_0} \tilde{\rho}_{\alpha_0} 
= \sum_{\alpha_0 , \beta_0 , \beta_1 } \delta_{\alpha_1 \beta_1} I_{\alpha_1 \alpha_1} I_{\beta_1 \beta_0} \delta_{\alpha_0 \beta_0} I_{\alpha_0 \alpha_0} \tilde{\rho}_{\alpha_0} \, .
\end{eqnarray}
This procedure is easily iterated to general values of $N$, the tensor network diagram for $N=3$ is depicted in Fig.~\ref{fig:tensordiag} (b). The evaluation of the path sum in Eq.~(\ref{path_I}) is therefore equivalent to the contraction of the corresponding tensor network. The $i-$th step of the tensor network contraction procedure consists of identifying the tensors with $k_1 = i-1$ as a matrix product state, which is applied to the matrix product operator consisting of the tensors with $k_1 = i$. This operation is done approximately by use of singular value decompositions where all singular values $\lambda < \epsilon \cdot \lambda_{\text{max}}$ are truncated where $\varepsilon$ is a convergence parameter and $\lambda_{\text{max}}$ is the largest singular value. Note that this operation is exact in the limit $\varepsilon \rightarrow 0$. This yields $\tilde{\rho}(t)$, which we use to compute $P(t) = \Braket{\sigma_z} (t) = \operatorname{tr} (\tilde{\rho}(t) \sigma_z)$.

\section{Convergence Analysis}
\begin{figure}[h]
\includegraphics[width=83mm]{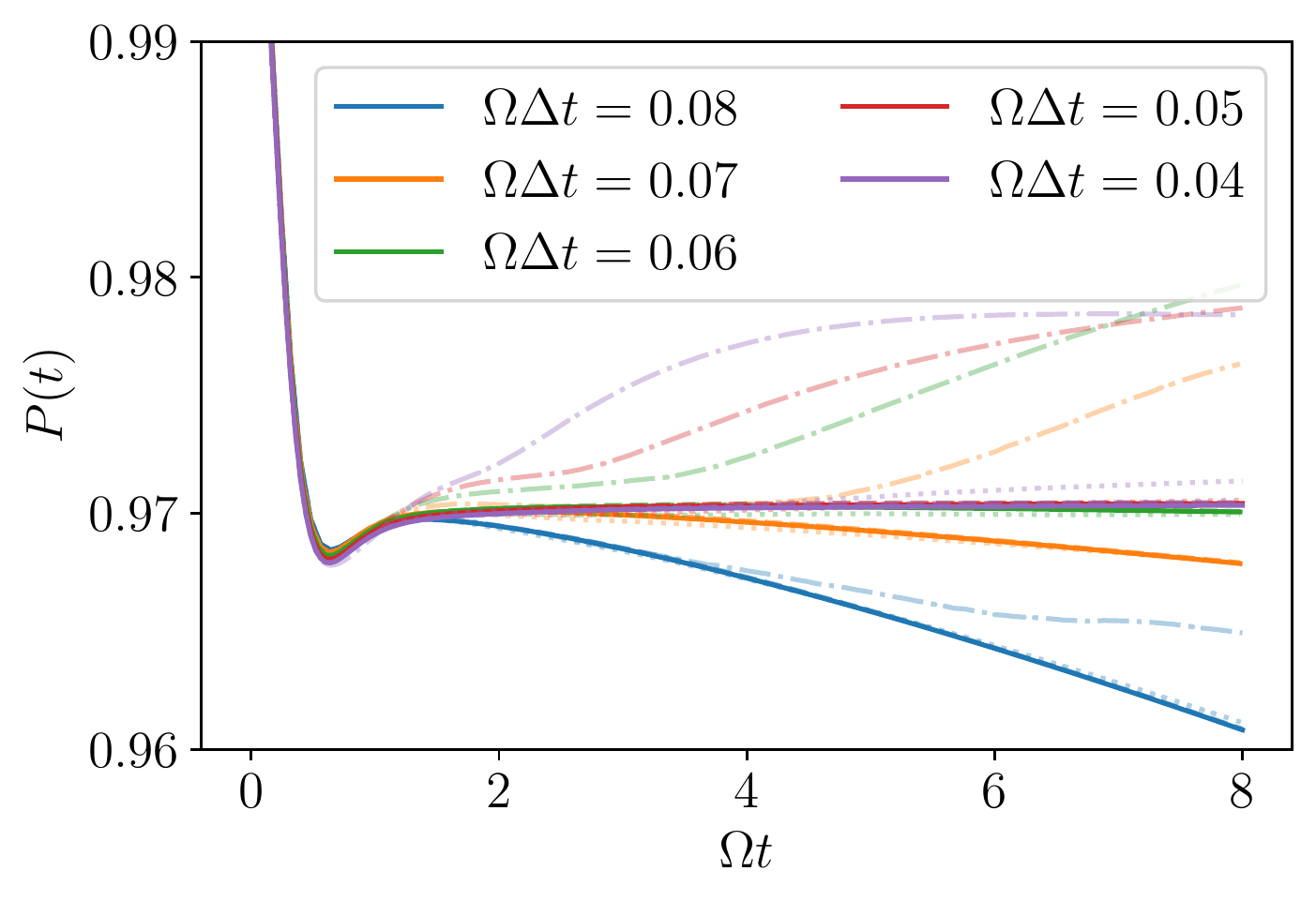}
\caption{Polarization $P(t)$ for $T=0$, $s = 0.3$, $\alpha = 0.8$, $\omega_c=10 \Omega$ and different values for the time discretization $\Delta t$. Dashdotted, dotted, dashed, and solid lines correspond to $\varepsilon = 10^{-5}$, $\varepsilon = 10^{-6}$, $\varepsilon = 10^{-7}$, and $\varepsilon = 10^{-8}$ respectively. Note that the dashed lines are only barely visible as they almost perfectly align with the solid lines.}
\label{fig:convergence_analysis}
\end{figure}

We have only made two approximations which are systematically controlled by the convergence parameters $\Delta t$ and $\varepsilon$. As smaller values of $\Delta t$ also require smaller values of $\varepsilon$, we need to converge the result for a given value of $\Delta t$ with respect to $\varepsilon$ first and then repeat for smaller values of $\Delta t$ until convergence is found. As an example, we consider the case of coupling strength $\alpha = 0.8$ and spectral exponent $s=0.3$ (see Fig.~\ref{fig:convergence_analysis}), which is among the most challenging cases studied in the main text. This is due to the strong coupling requiring a fine time discretization and the low spectral exponent causing long bath memory, which causes singular values to fall off relatively slowly. Note that for all of the values of $\Delta t$ considered in Fig.~\ref{fig:convergence_analysis}, we achieve excellent convergence with respect to $\varepsilon$ and that the dynamics is also converged with respect to the time discretization.

\section{Polarized Initial Condition}
As in Ref.~[16], we next consider polarized initial conditions where the bath is equilibrated to a spin with $\Braket{\hat{\sigma}_z} (t=0) = 1$, i.e., $\rho_{B}(0) \propto \exp{ \left\lbrace- \left( \hat{H}_B + \hat{H}_{\text{int}} (\hat{\sigma}_z = 1) \right) / T \right\rbrace }$ as opposed to the unpolarized initial condition $\rho_{B}(0) \propto \exp{ \left\lbrace-  \hat{H}_B / T   \right\rbrace }$ used in the main text. Following the derivation of Ref.~[15], the polarized initial condition leads to the additional factors in the TEMPO scheme 
\begin{equation}
    \prod_{k>0} \tilde{F}_{s_{k}^\pm s_{0}^\pm}^{(k 0)} = \prod_{k>0} \exp  \left[ \frac{i}{2} (s_k^+ - s_k^- ) \tilde{\eta}_{k0} \right]
\end{equation}
in Eq.~(\ref{red}), where
\begin{equation}
    \tilde{\eta}_{k0} = \int_0^\infty  \frac{2J(\omega)}{\omega^2} \left[ \sin{\left(\omega k \Delta t \right)} - \sin{\left( \omega (k-1) \Delta t \right) }\right] \mathrm{d}\omega.
\end{equation}
This can be straightforwardly included in the time-evolving matrix product operator technique presented in Section~\ref{tempo} as the structure of Eq.~(\ref{red}) remains unchanged.

\begin{figure}[h]
\centering
\begin{tabular}{cc}
\includegraphics[width=0.50\textwidth]{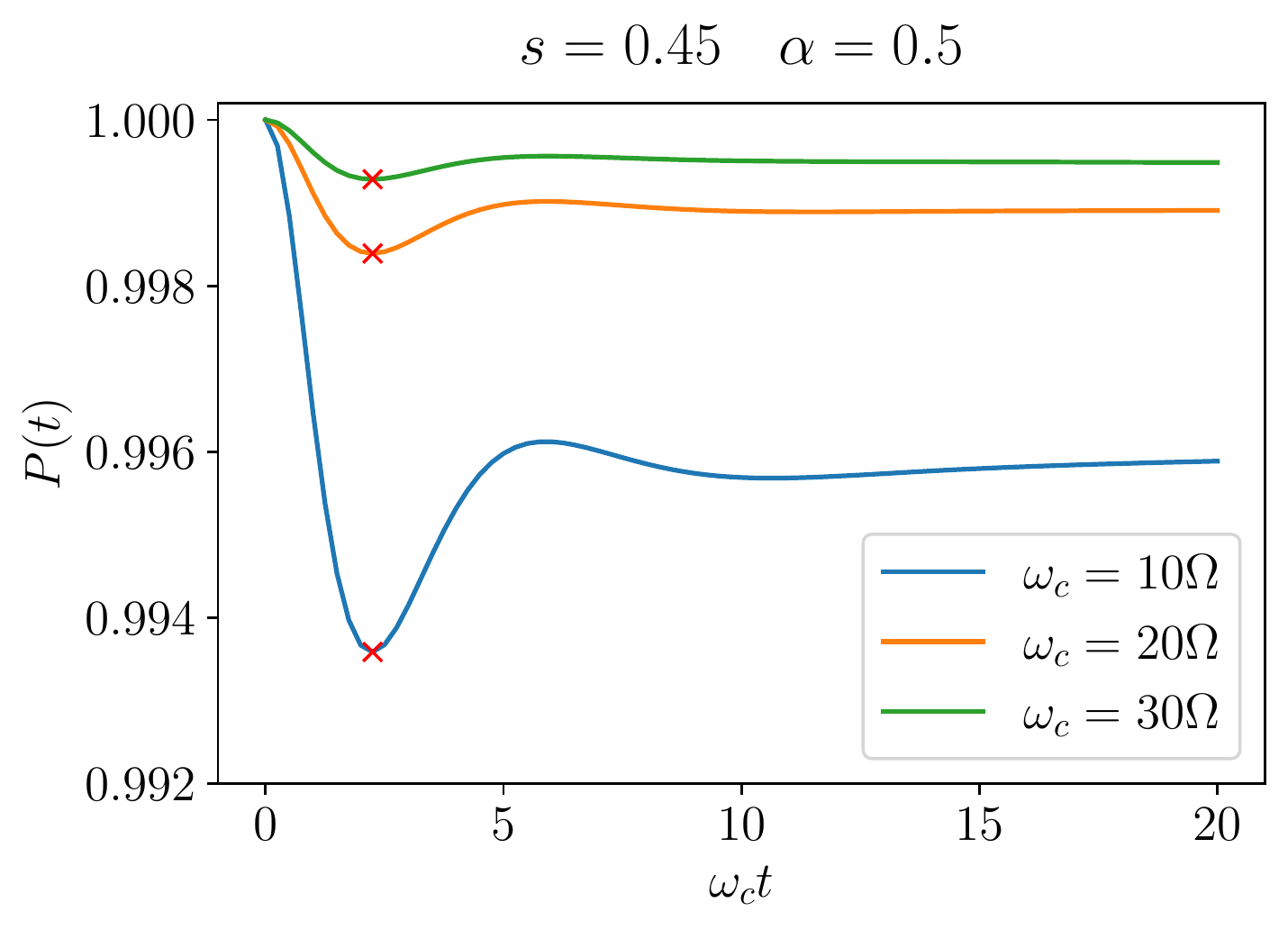}&
\includegraphics[width=0.50\textwidth]{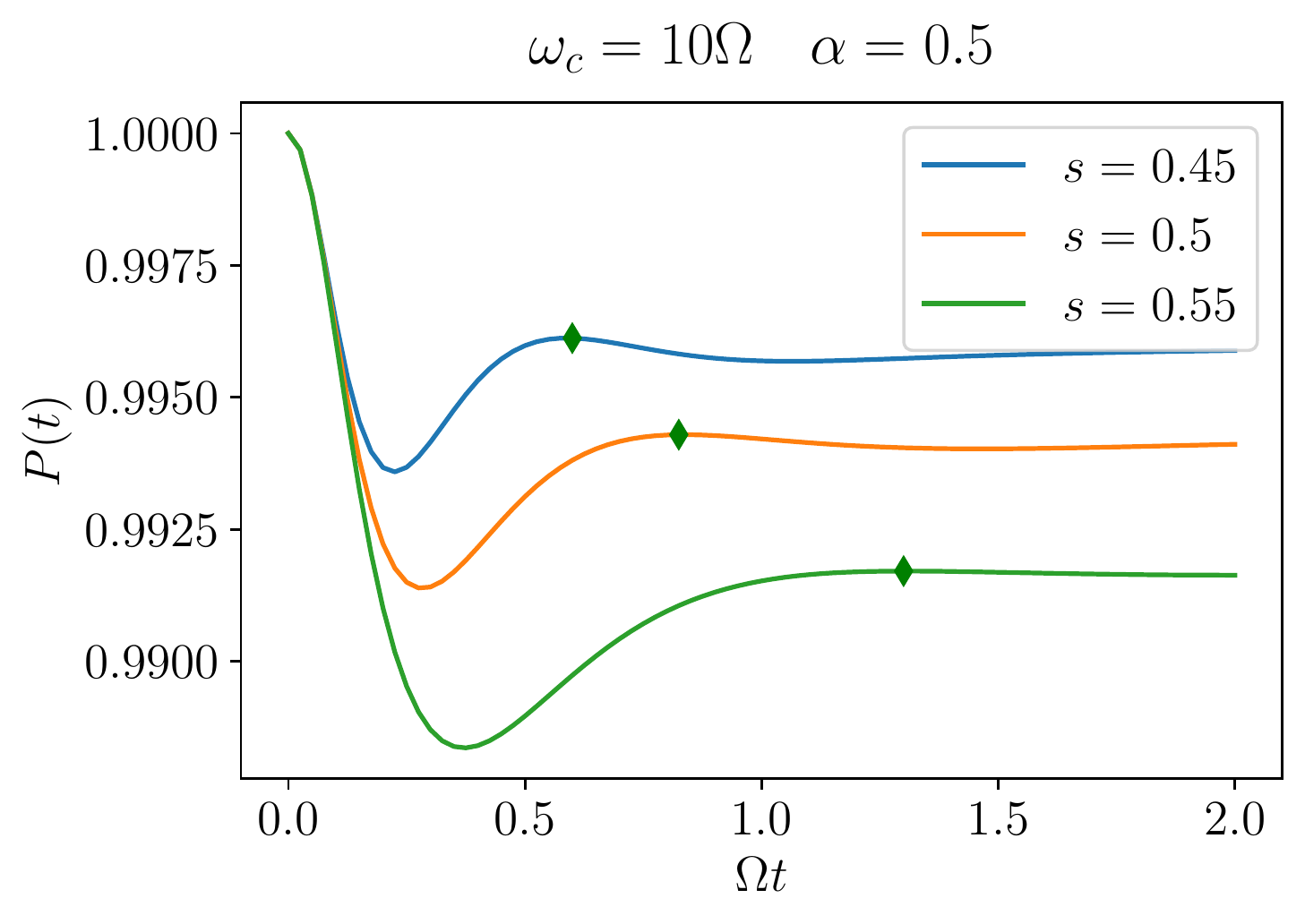} \\
 $\qquad \quad$ \textbf{(a)}   & $\qquad \qquad \quad$ \textbf{(b)}     \\
\end{tabular}
\caption{ Polarization $P(t)$ with polarized initial conditions for $T=0$, $\alpha = 0.5$ and (a) for  $s=0.45$, and different bath cut-off frequencies $\omega_c$ as indicated (local minima are marked with a red cross), and (b) for $\omega_c = 10 \Omega $, and different spectral exponents $s$ (local maxima are marked with a green diamond).}
\label{fig:dyn}
\end{figure}

In Fig.~\ref{fig:dyn} (a), we show that also for the polarized initial condition, the minimum of $P(t)$ occurs at times $t_{\rm min}\simeq \text{const.}/\omega_c $ (notice the scaling on the time axis). While in the unpolarized case the renormalization starts to dominate the oscillatory frequency at a coupling strength when the damping rate roughly equals the frequency, the polarized initial condition causes a strong frequency renormalization already at weaker coupling (see discussion in Ref.~[16]). Therefore,  maxima are observed also in the pseudo-coherent phase. For the polarized case it is thus less obvious when the dynamics becomes fully governed by the bath. Nevertheless, the bath dominates the dynamics as becomes clear from Fig.~\ref{fig:dyn} (a) here and the estimate of the oscillation frequency in Ref.\ [16] in the limit $s \rightarrow 0$. Therefore, for both initial bath conditions at weakest coupling strengths and $s \lesssim 0.5$ there is damped oscillatory behavior with an oscillation frequency determined by the system, i.e., by $\Omega$. When increasing the coupling strength, the oscillation frequency becomes completely determined by the bath. This behavior is characteristic for the pseudo-coherent phase. The observed oscillatory dynamics is thus not rooted in the system, but is the result of the system being slaved to the bath. 

Furthermore, we do not find a qualitative change in the dynamical behavior at $s=0.5$ (see Fig.~\ref{fig:dyn} (b)) for large coupling strength. This behavior of the pseudo-coherent phase extends to all $0 < s < 1$ at strong coupling and reflects bath dynamics due to finite $\omega_c$ to which the system is fully slaved. Specifically, when entering the pseudo-coherent phase with increasing coupling for a fixed s with $0.5 < s < 1$ where oscillatory behavior re-emerges after the dynamics was fully incoherent for a smaller coupling, it becomes clear that the pseudo-coherent phase is distinct from the conventional damped coherent dynamics at weak coupling – independent of the choice of initial condition of the bath.
